\documentclass[%
 aip,
pof,
 amsmath,amssymb,
preprint,%
]{revtex4-1}

\usepackage {CJK}

\usepackage{graphicx}
\usepackage{dcolumn}
\usepackage{bm}

\usepackage[utf8]{inputenc}
\usepackage[T1]{fontenc}
\usepackage{mathptmx}

\begin{document}

\begin{CJK*}{UTF8}{} 

\preprint{Preprint submit to \emph{Physics of Fluids}}

\title{Migration of self-propelling agent in a turbulent environment with minimal energy consumption}

\author{Ao Xu}
\email{Corresponding author: axu@nwpu.edu.cn (Ao Xu)}
 \affiliation{School of Aeronautics, Northwestern Polytechnical University, Xi'an 710072, China}%
  \affiliation{Institute of Extreme Mechanics, Northwestern Polytechnical University, Xi'an 710072, China}%

\author{Hua-Lin Wu}
 \affiliation{School of Aeronautics, Northwestern Polytechnical University, Xi'an 710072, China}%

\author{Heng-Dong Xi}
 \affiliation{School of Aeronautics, Northwestern Polytechnical University, Xi'an 710072, China}%
 \affiliation{Institute of Extreme Mechanics, Northwestern Polytechnical University, Xi'an 710072, China}%

\date{\today}

\begin{abstract}
We present a numerical study of training a self-propelling agent to migrate in the unsteady flow environment.
We control the agent to utilize the background flow structure by adopting the reinforcement learning algorithm to minimize energy consumption.
We considered the agent migrating in two types of flows: one is simple periodical double-gyre flow as a proof-of-concept example,
while the other is complex turbulent Rayleigh-B\'enard convection as a paradigm for migrating in the convective atmosphere or the ocean.
The results show that the smart agent in both flows can learn to migrate from one position to another while utilizing background flow currents as much as possible to minimize the energy consumption,
which is evident by comparing the smart agent with a naive agent that moves straight from the origin to the destination.
In addition, we found that compared to the double-gyre flow,
the flow field in the turbulent Rayleigh-B\'enard convection exhibits more substantial fluctuations, and the training agent is more likely to explore different migration strategies;
thus, the training process is more difficult to converge.
Nevertheless, we can still identify an energy-efficient trajectory that corresponds to the strategy with the highest reward received by the agent.
These results have important implications for many migration problems such as unmanned aerial vehicles flying in a turbulent convective environment, where planning energy-efficient trajectories are often involved.
\footnote{
This article may be downloaded for personal use only.
Any other use requires prior permission of the author and AIP Publishing.
This article appeared in Xu \emph{et al.}, Phys. Fluids \textbf{34}, 035117 (2022) and may be found at \url{https://doi.org/10.1063/5.0082845}.
}
\end{abstract}

\maketitle
\end{CJK*}

\section{\label{Section1}Introduction}
Soaring birds and gliders often use spatially and temporally localized warm rising atmospheric currents to stay aloft and fly higher. \cite{cone1962thermal}
The so-called \emph{thermal soaring} behavior  \cite{ludlam1953reviews} can save a vast amount of energy by minimizing the flapping of wings for birds or motor power supplies for gliders over long distances. \cite{bencatel2013atmospheric}
For example, the Andean condor can even soar over 5h (covering about 172 km) without flapping. \cite{williams2020physical}
However, the air currents in the troposphere are turbulent; thus, using wobbly gusts of air to stay airborne has not always been a simple task.
Ideal conditions for thermal soaring typically occur when a strong temperature gradient between the surface of the Earth and the top of the atmospheric boundary layer creates convective thermals.
The convective thermals exhibit general turbulence characterized by strongly fluctuating flow velocities.

Learning about details of thermal soaring can improve understanding of the main features of flight trajectories and optimization strategies.
In 1958, MacCready \cite{maccready1958optimum} proposed a theory on flight optimization
and gave a gliding polar curve (the relationship between horizontal speed and the corresponding vertical one)  to calculate the best slope to take before an upcoming thermal.
Since then, gliders have tried to adjust their gliding speed to the expected thermal climb rate according to their polar curve. With the aid of measured polar curves,
Akos \emph{et al.} \cite{akos2008comparing} found that there are relevant common features in the way that falcons and the world's leading paraglider pilots use thermals,
which are also close to the optimal soring strategy predicted by MacCready's theory.
To apply the above soaring strategy for an unmanned aerial vehicle (UAV) to take advantage of thermals,
Allen and Lin \cite{allen2007guidance} adopted autonomous soaring algorithms to detect and exploit thermals.
He used the aircraft's total energy state to detect and soar within thermals and the estimated thermal size and position to calculate guidance commands for soaring flight.
On the other hand, reinforcement learning (RL) methods are promising to deliver effective strategies of soaring flight.
For example, Wharington and Herszberg \cite{wharington1998control} used a neural-based algorithm to locate the thermal core.
However, they only considered the learning problem of finding the center of a stationary thermal without turbulence.
Later, Akos \emph{et al.} \cite{akos2010thermal} demonstrated that such simple rules would fail in the presence of velocity fluctuations.
Thus, soaring strategies that could work in real turbulent work are urgently needed.
To identify effective soaring strategies of flight in turbulent flows, Reddy \emph{et al.} \cite{reddy2016learning,reddy2018glider} used reinforcement learning algorithms to train gliders to travel through complex choppy air currents.
Environment cues, such as an increase in the twisting force of the wind that indicates rising air, can be sensed.
However, in the work of Reddy \emph{et al.}, their motivation is to train the glider to employ spiraling patterns to ascend higher in regions of strong upwelling currents.
In practical flight tasks, we would also expect the UAV to fly from one position to another
while utilizing thermal soaring as much as possible to extend the flight duration and reduce the energy consumption. \cite{dbouk2021quadcopter}

Previously, reinforcement learning algorithms have also been applied to navigate micro gravitactic swimmers
to escape local fluid traps and reach the highest altitude, \cite{colabrese2017flow, gustavsson2017finding}
to accumulate in regions of intense negative vorticity, \cite{colabrese2018smart}
and to reach a target position with minimal time. \cite{biferale2019zermelo,alageshan2020machine,muinos2021reinforcement}
Shape effects (e.g., elliptical or ellipsoidal shape) of asymmetric swimmers have also been considered with the goal of moving upwards. \cite{qiu2020swimming,qiu2022navigation}
In this work, our motivation is to train an active self-propelling agent to migrate from one position to another with minimal energy cost.
This motivation stems from reducing the energy consumption for UAVs during their flight.
The rest of this paper is organized as follows.
In Sec. \ref{Section2}, we present details for the reinforcement learning algorithm to train the self-propelling agent to find an energy-efficient trajectory.
In Sec. \ref{Section3}, we first train the agent in the unsteady double-gyre flow, \cite{shadden2005definition} which is a simple periodic flow environment with an analytical solution for the flow field,
as a proof-of-concept example.
In Sec. \ref{Section4}, we then train the agent in the turbulent Rayleigh-B\'enard convection, \cite{xia2013current}
which is a much more complex turbulent environment with a strongly fluctuating flow field as a paradigm for migration in the atmosphere or the ocean.
In Sec. \ref{Section5}, the main findings of this work are summarized.

\section{\label{Section2}Dynamics of the self-propelling agent}
\subsection{Optimal control via the reinforcement learning algorithm}

We aim to plan an energy-efficient trajectory for the self-propelling agent in the unsteady flow environment.
Traditional approaches, such as the optimal navigation theory,  \cite{callier2012linear,pontryagin1987mathematical} may be sensitive to small disturbances in the chaotic system.
Thus, we adopt the emerging reinforcement learning (RL) algorithm to optimize the trajectory in this work. \cite{mehta2019high,carleo2019machine,brunton2020machine,garnier2021review}
In the RL algorithm, the agent observes the state of the environment and then decides on an action to take.
If the agent receives a reward (or a penalty) for that action, it is more likely to repeat (or forego) the action in the future.
Overall, the agent learns by trial and error and eventually achieves its goal. \cite{sutton2018reinforcement,cichos2020machine}
In the double-gyre flow, the observation variables include the background flow velocity $\mathbf{u}_{\text{fluid}}$, the agent's spatial coordinates $\mathbf{x}_{\text{agent}}$, and the current time $t$;
in the turbulent Rayleigh-B\'enard convection, we also include the fluid temperature $T$ in addition to the above-mentioned observation variables.
The action variable is that the agent generates propelling velocity of $\mathbf{u}_{\text{propel}}$.

The model-free reinforcement learning algorithms can generally be classified into two categories:
one is the policy optimization method, in which the parameter $\theta$  is optimized to maximize the performance objective $J(\pi_{\theta})$,
and the other is the Q-learning method, in which the agent takes action $a$ that tried to maximize the optimal action-value function, i.e., $a(s)=\arg \max \limits_{a}Q_{\theta}(s,a)$.
Here, $s$ denotes the state of the environment,  $\pi_{\theta}$ denotes the parameterized stochastic policy, and $Q_{\theta}(s,a)$ approximates the optimal action-value function  $Q^{*}(s,a)$.
However, the policy optimization method is inefficient in sampling, because it cannot reuse data to train the model,
while the Q-learning method tends to be less stable because it indirectly optimizes the agent performance.  \cite{tsitsiklis1997analysis,szepesvari2010algorithms}
A trade-off between these two methods is the soft actor-critic (SAC) method, \cite{haarnoja2018soft}
in which the actor aims to maximize the expected reward (i.e., succeed at the task) while also maximizing entropy (i.e., acting as randomly as possible).
In entropy-regularized reinforcement learning, the optimization problem can be described as
\begin{equation}
\pi^{*}(\theta)=\arg \max \limits_{\pi} \mathop{E} \limits_{\tau \sim \pi}\left[ \sum_{t=0}^{\infty}\left(R(s_{t},a_{t},s_{t+1})+\alpha H(\pi(\cdot | s_{t})) \right) \right]
\end{equation}
In the above equation, $\pi^{*}$ is the optimal policy.
The reward function $R$  depends on the current state of the environment $s_{t}$, the action just taken $a_{t}$, and the next state of the environment $s_{t+1}$.
$\alpha$  is the trade-off coefficient.
The entropy $H$ of  $\tau$ is computed from its distribution $\pi$  as $H(\pi(\cdot|s_{t}))=\mathop{E} \limits_{\tau \sim \pi}\left[ -\log \pi(\tau)\right]$ .
More details on the SAC method can be found in Ref.~\cite{haarnoja2018soft}.

In this work, we assume the rewards gained by the agent are simultaneously affected by
its current state, energy consumption, and time consumption.
We design the reward function as
\begin{equation}
r(t)=r_{s}(t)+r_{e}(t)+r_{h}(t)
\end{equation}
Here, $r_{s}$ denotes the reward contributed by the current state of the agent.
We assume that if the agent migrates out of the flow domain, it will receive a penalty of -10;
if the agent is getting closer to the destination, it will receive a basic reward of $e_{\text{basic}}$.
Thus, we can express  $r_{s}$ as
\begin{equation}
r_{s}=\left\{
    \begin{aligned}
& -10, \ &\text{agent is out of the flow domain} \\
& e_{\text{basic}}, \ &\left\|\mathbf{x}_{\text{agent}}^{t+1}-\mathbf{x}_{\text{goal}}\right\|_{2}<\left\|\mathbf{x}_{\text{agent}}^{t}-\mathbf{x}_{\text{goal}}\right\|_{2} \\
& 0, \  &\text{otherwise}
    \end{aligned}
\right.
\end{equation}
Here, $r_{e}$ denotes the reward contributed by the energy consumption of the agent.
We assume that if the propelling velocity of the agent $\mathbf{u}_{\text{propel}}$ is in alignment with that of the background flow $\mathbf{u}_{\text{fluid}}$,
namely, the angle between these two vectors is $\alpha \le 90^{\circ}$, the agent will receive a reward of  $e_{\text{basic}}+(e_{\text{max}}-e)$,
where $e=0.5\left\|\mathbf{u}_{\text{propel}}\right\|^{2}$ and  $e_{\text{basic}}=e_{\max}=0.5(\left\|\mathbf{u}_{\text{propel}}\right\|)^{2}_{\max}$,
suggesting that when the agent migrates, it follows the background flow direction,
the higher the agent generates propelling velocity, the lower the reward it receives;
otherwise, it will receive a penalty of  $-(e_{\text{basic}}+e)$,
suggesting that when the agent migrates against the background flow direction,
the higher the agent generates propelling velocity, the higher the penalty it receives.
Thus, we can express  $r_{e}$ as
\begin{equation}
r_{e}=\left\{
    \begin{aligned}
& e_{\text{basic}}+(e_{\max}-e), \ & \text{for} \ 0^{\circ}\le \alpha \le 90^{\circ}  \\
& -(e_{\text{basic}}+e), \ & \text{for} \ 90^{\circ}< \alpha \le 180^{\circ}
    \end{aligned}
\right.
\end{equation}
Here, $r_{h}$ denotes the reward contributed by the time consumption of the agent.
We assume that if the agent cannot reach the destination within a maximum time of  $t_{\max}$, it will receive a penalty of -5;
if the agent is within $\delta_{0}$ from the destination (here, $\delta_{0}$ denotes a small threshold value),
we assume the agent reaches the destination and it will receive a reward that is inversely proportional to the time taken (with coefficient $\epsilon$) during the migration,
suggesting the sooner the agent reaches the destination, the higher the reward it receives.
Thus, we can express  $h_{t}$ as
\begin{equation}
r_{h}=\left\{
    \begin{aligned}
& -5, \ &  t\ge t_{\max}  \\
& \varepsilon(t_{\max}-t), \ & \left\|\mathbf{x}_{\text{agent}}^{t}-\mathbf{x}_{\text{goal}}\right\|_{2}<\delta_{0} \\
& 0, \ &  \text{otherwise}  \\
    \end{aligned}
\right.
\end{equation}
Our specially designed reward function also implies that when the agent migrates toward the destination
(i.e., $\left\|\mathbf{x}_{\text{agent}}^{t+1}-\mathbf{x}_{\text{goal}}\right\|_{2}<\left\|\mathbf{x}_{\text{agent}}^{t}-\mathbf{x}_{\text{goal}}\right\|_{2}$)
following the background flow direction (i.e.,  $0^{\circ}\le \alpha \le 90^{\circ}$),
in each timestep, reducing energy consumption will be the agent's primary objective, while approaching the destination will be its second objective [because $e_{\text{basic}}+(e_{\max}-e)>e_{\text{basic}}$].
In addition, if the agent has to reach the destination within a short time (i.e.,  $t_{\max}$ is not long enough for the agent to freely explore the environment),
in each episode, approaching the destination will be the agent's primary objective, while reducing energy consumption will be its second objective
[because $E\left[\sum r_{h} \right]>E\left[\sum (r_{s}+r_{e})\right]$].

\subsection{Kinematic model for the self-propelling agent}
We restricted the maximum propelling velocity of the agent $\mathbf{u}_{\text{propel}}$
to be less than the largest background flow velocity,
such that intelligent planning of the agent can well utilize the background flow structure,
also to mimic the limited propulsion available in real-world scenarios.
We assume that, without control, the velocity of the agent equals the velocity of the background fluid flow $\mathbf{u}_{\text{fluid}}$;
meanwhile, the agent can take action to generate its own relative velocity $\mathbf{u}_{\text{propel}}$.
Then, with control, we can model the agent's velocity in the unsteady flow as
$\mathbf{u}_{\text{agent}}=\mathbf{u}_{\text{fluid}}+\mathbf{u}_{\text{propel}}$.
The position of the agent is updated via the relation  $d\mathbf{x}_{\text{agent}}/dt=\mathbf{u}_{\text{agent}}$.
It is worth mentioning that the present kinematic model for the agent is far from a realistic one in the industry.
Here, we adopt this simple kinematic model to disentangle the coupling between the chaotic flow of the carrier fluid and the complex motion of the agent.
On the other hand,
a more complex kinematic model for the self-propelling agent
that includes inertial and rotational dynamics, \cite{zhang2008optimal}
flapping motion, \cite{wang2021optimal,liu2021full}
or even the flexible motion of the propelling agent \cite{wang2020optimal,yu2021collective,liu2020propulsive} can be considered in the future work.

\section{\label{Section3}Migration in the unsteady double-gyre flow}
\subsection{Numerical simulation of unsteady double-gyre flow}
The double-gyre flow field has the analytical description of the velocity field.
It has been used to study mixing and coherent structures in large-scale ocean circulation.
The flow is defined on a nondimensional domain of  $[0,2]\times[0,1]$.
The double-gyre velocity field is derived from the stream function
\begin{equation}
  \phi(x,y,t)=A \sin \left[ \pi f(x,t) \right] \sin(\pi y)
\end{equation}
and the resulting velocity field is
\begin{subequations}
\begin{align}
&   u(x,y,t)=-\pi A \sin \left[ \pi f(x,t) \right] \cos(\pi y) \\
&   v(x,y,t)=-\pi A \cos \left[ \pi f(x,t) \right] \sin(\pi y)
\end{align}
\end{subequations}
In the above equations, the time dependency is introduced by
\begin{equation}
  f(x,t)=a(t)x^{2}+b(t)x
\end{equation}
with time-dependent coefficient
\begin{equation}
  a(t)=\varepsilon \sin(\omega t), \ \ \ b(t)=1-2\varepsilon \sin(\omega t)
\end{equation}
Here, $A$ determines the magnitude of the velocity vectors, $\varepsilon$ is the amplitude of the motion of the separation point on the $x$-axis and $\omega$ is the angular oscillation frequency.
Unless otherwise mentioned, we adopt the parameter sets of  $A=0.1$, $\varepsilon=0.25$,  and  $\omega=2\pi / 10$ as default values. \cite{shadden2005definition}

\subsection{Training results and discussion}
\begin{figure}
  \centering
  \includegraphics[width=12cm]{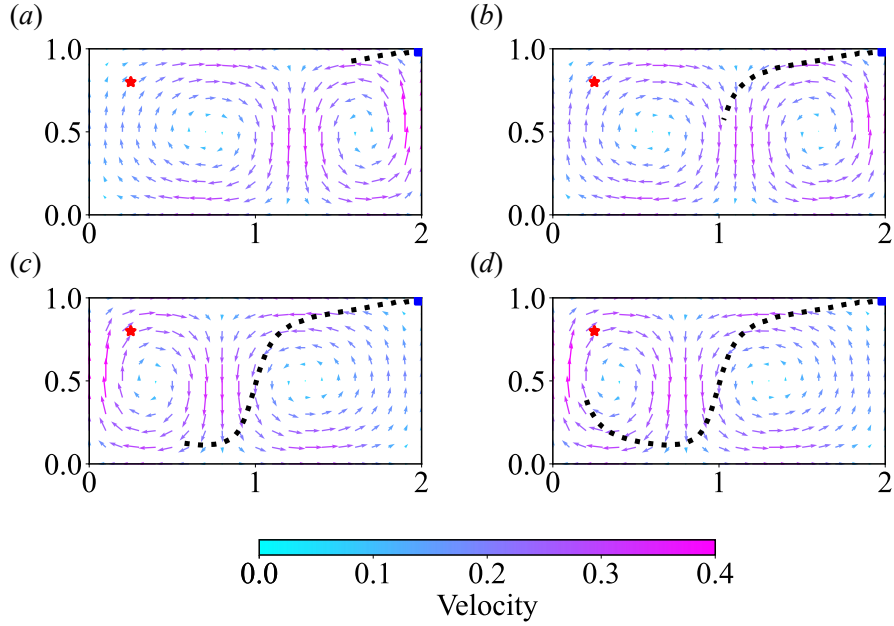}
  \caption{\label{Fig-DG_trajectory} Trajectory (black dotted line) of the smart agent in the double-gyre flow at
  (\textit{a})  $t = 2.2$, (\textit{b})  $t = 4.2$, (\textit{c})  $t = 6.7$, (\textit{d})  $t = 8.2$.
  The vectors denote the velocity field of the double-gyre flow, and they are color-coded by the velocity magnitude.
  (Multimedia view)}
\end{figure}

In training, we restrict the maximum propelling velocity of the agent along either $x$- or $y$-direction to be less than $A=0.1$,
which leads to  $\left\|\mathbf{u}_{\text{propel}}\right\|^{2}\le 0.02$.
We show the instantaneous trajectories for the smart agent in the double-gyre flow in Fig. \ref{Fig-DG_trajectory} (Multimedia view).
We released the agent at (2.0, 1.0) position, namely the top-right corner in the domain (marked by the blue square in the plot).
The agent's goal is to reach the destination position of (0.25, 0.8), marked by the red star in the plot, with minimal energy consumption.
Here, we assume the agent reaches the destination so long as its position is within $\delta_{0}=0.05$ from (0.25, 0.8).
Initially, the smart agent will move leftward to utilize the horizontal currents in the top-right region [see Fig. \ref{Fig-DG_trajectory}(a)].
When the smart agent reaches the top-left corner of the right gyre at the position around (1, 0.9),
it will move downward to utilize the vertical currents [see Fig. \ref{Fig-DG_trajectory}(b)].
After that, it will drift into the bottom-right corner of the left gyre at the position around (1, 0.1) [see Fig. \ref{Fig-DG_trajectory}(c)]
and then follow the horizontal leftward and vertical upward current to reach the destination [see Fig. \ref{Fig-DG_trajectory}(d)].
The smart agent follows background currents as much as possible, except at the ridge between the two gyres around $x = 1$,
where the agent will consume more energy and drift from the right gyre into the left gyre.
Because the structure of the flow leaves its mark on the trajectories of the agents carried by turbulent flows, \cite{laurent2021turbulence}
to quantitatively describe the relationship between the orientation of the propelling velocity vector (i.e., $\theta_{\text{propel}}$)
and the orientation of the fluid velocity vector (i.e.,  $\theta_{\text{fluid}}$),
we calculate their cross-correlation coefficient as
\begin{equation}
C=\langle \left[\theta_{\text{propel}}(t)-\langle\theta_{\text{propel}} \rangle \right]\left[\theta_{\text{fluid}}(t)-\langle\theta_{\text{fluid}} \rangle \right] \rangle/(\sigma_{\theta_{\text{propel}}}\sigma_{\theta_{\text{fluid}}})
\end{equation}

The resulting correlation coefficient of 0.60 suggests
that the orientations of the propelling velocity vector and fluid velocity vector are statistically relevant.
We also visualize the propelling velocity vector and fluid velocity vector of the smart agent in the Appendix.

\begin{figure}
  \centering
  \includegraphics[width=12cm]{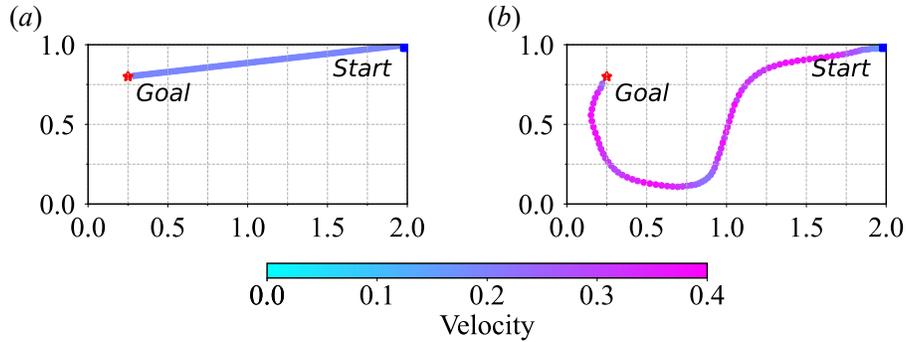}
  \caption{\label{Fig-DG-trajectory-compare}
  Comparison of trajectories in the double-gyre flow for (\textit{a}) a naive agent moving straightly and (\textit{b}) a smart agent utilizing the flow structure to save energy.
  The trajectories are color-coded by the instantaneous velocity magnitude.}
\end{figure}

To demonstrate that the smart agent can, indeed, reduce energy consumption, we compare the strategy of the smart agent with the naive agent.
Here, the naive agent moves straight from the origin to the destination with constant velocity, which may be the "simplest" way for an agent to migrate from one position to another.
To make a fair comparison on the energy consumption, we set the naive agent spending the same amount of time $t_{\text{total}}$ as the smart agent migrating from the origin to the destination.
Thus, the velocity magnitude of the agent is $\|\mathbf{u}_{\text{agent}}\|=\|\mathbf{x}_{\text{goal}}-\mathbf{x}_{\text{start}} \|/t_{\text{total}}$ and its direction point from the origin to the destination.
In Fig. \ref{Fig-DG-trajectory-compare}, we show the trajectories of the naive agent and the smart agent, which are color-coded by the instantaneous velocity magnitude.
We can see from Fig. \ref{Fig-DG-trajectory-compare} that the naive agent migrates slower than the smart agent,
because the naive agent's travel distance is shorter than that of the smart agent.
Although the smart agent migrates faster, it does not indicate that the smart agent will consume more energy because the smart agent can utilize the flow currents to save energy.

\begin{figure}
  \centering
  \includegraphics[width=12cm]{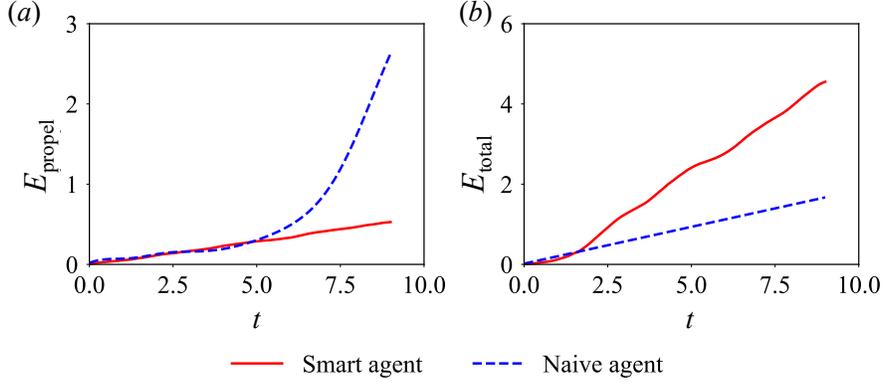}
  \caption{\label{Fig-DG-energy-compare}
  Comparison of the accumulative (\textit{a}) energy consumption $E_{\text{propel}}$ and (\textit{b}) total kinetic energy $E_{\text{total}}$ of the smart agent and the naive agent in the double-gyre flow.}
\end{figure}

To make a quantitative comparison, we plot the time series of accumulative energy consumed by the agents, which is calculated as
\begin{equation}
E_{\text{propel}}(t)=\int_{0}^{t}\frac{1}{2}\left\|\mathbf{u}_{\text{propel}}(\tau)\right\|^{2}d\tau
\end{equation}
In this work, we assume the agent generates its own velocity of  $\mathbf{u}_{\text{propel}}$, which is responsible for its energy consumption.
As shown in Fig. \ref{Fig-DG-energy-compare}(a), we can see that both agents consume almost the same amount of energy during the initial period (i.e., around  $t<6$),
which is due to similar trajectories in the flow field (see Fig. \ref{Fig-DG-trajectory-compare}), and both agents migrate in the top area of the right gyre.
After that (i.e., around $t>6$), the smart agent moves downward to utilize the flow currents,
while the naive agent continues to move straight toward the destination.
Crossing the ridge of the gyre and migrating reversely against the flow currents will require substantial energy consumption,
as evident from the much higher energy consumption at $t<6$  for the naive agent [see Fig. \ref{Fig-DG-energy-compare}(a)].
We also compare the accumulative total kinetic energy of the agents [see Fig. \ref{Fig-DG-energy-compare}(b)], which is calculated as
\begin{equation}
E_{\text{total}}(t)=\int_{0}^{t}\frac{1}{2}\left\|\mathbf{u}_{\text{agent}}(\tau)\right\|^{2}d\tau
\end{equation}
After reaching the destination, the accumulative total kinetic energy of the smart agent is more than twice that of the naive agent,
while the smart agent only consumed almost one-fifth of the energy,
suggesting the smart agent can efficiently utilize the energy provided by the background flow.

\begin{figure}
  \centering
  \includegraphics[width=17cm]{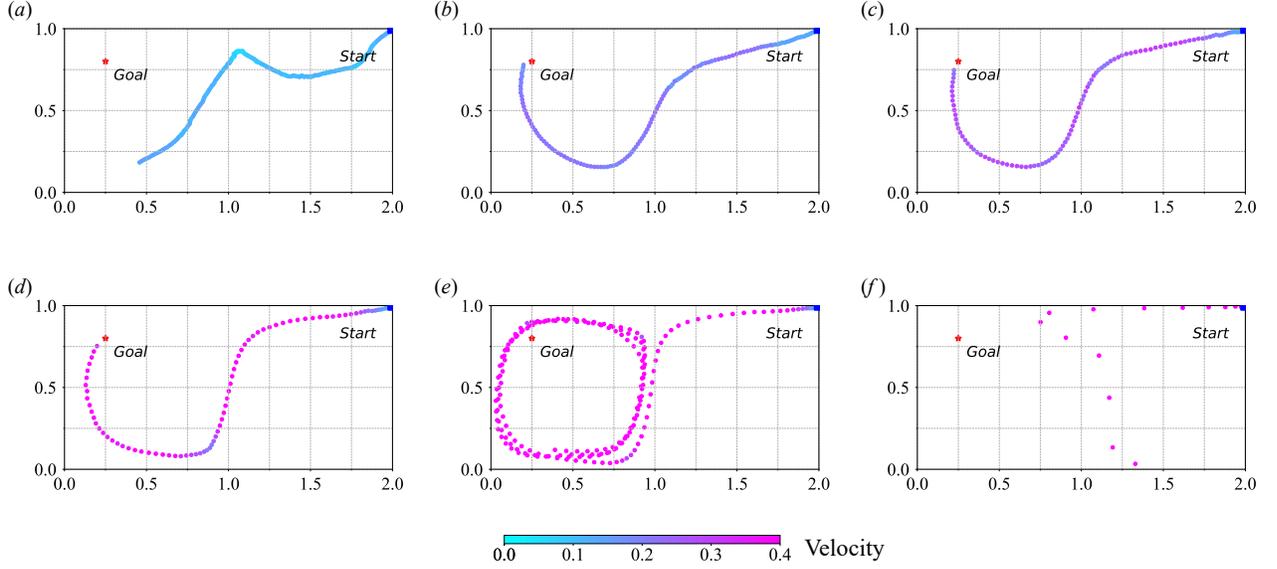}
  \caption{\label{Fig-DG-variousA}
  Trajectories for the smart agent in the double-gyre flow with various control parameters of the flow field: (\textit{a}) $A = 0.01$, (\textit{b}) $A = 0.05$, (\textit{c}) $A = 0.08$, (\textit{d}) $A = 0.12$, (\textit{e}) $A = 0.2$, and (\textit{f}) $A = 1.0$. The trajectories are color-coded by instantaneous velocity magnitude.}
\end{figure}

In the following, we test the robustness of the energy-efficient strategies with respect to the flow control parameters.
We varied $\varepsilon$ in the range of $0.025 \le \varepsilon \le 2.5$ while keeping $A$ and $\omega$  fixed as the default values.
The resulting optimized trajectories show minor differences, suggesting energy-efficient strategy is robust with the changes in the magnitude of oscillation in the $x$-direction.
Similarly, we varied  $\omega$ in the range of $0.2\pi/10 \le \omega \le 2\pi$  while keeping $A$ and $\varepsilon$  fixed as the default values.
The resulting optimized trajectories show minor differences, suggesting energy-efficient strategy is also robust with the changes in the angular oscillation frequency.
We then vary $A$ in the range of $0.01 \le A \le 1$  while keeping $\varepsilon$ and $\omega$  fixed as the default values.
As shown in Fig. \ref{Fig-DG-variousA}, the resulting optimized trajectories show significant differences:
when the carrier fluid flows at a low speed [see Figs. \ref{Fig-DG-variousA}(a) and \ref{Fig-DG-variousA}(b), for $A=0.01$ and $A=0.05$, respectively], the migration of the agent mostly depends on its own propulsion, and it slowly approaches the destination or even cannot reach the destination;
when the carrier fluid flows at a high speed [see Figs. \ref{Fig-DG-variousA}(e) and \ref{Fig-DG-variousA}(f), for $A=0.2$ and $A=1.0$, respectively], the migration of the agent is significantly influenced by the carrier flow, and it may miss the destination or even drift out the flow domain.
We also slightly increase or decrease 20\% of the $A$ value [see Figs. \ref{Fig-DG-variousA}(c) and \ref{Fig-DG-variousA}(d), for $A=0.08$ and $A=0.12$, respectively], and the resulting optimized trajectories show minor differences.
Thus, the optimized energy-efficient strategy is not sensitive to small perturbation of the magnitude of the velocity vectors, but it certainly changes when the magnitude of the velocity vectors varies in a larger range.

\begin{figure}
  \centering
  \includegraphics[width=12cm]{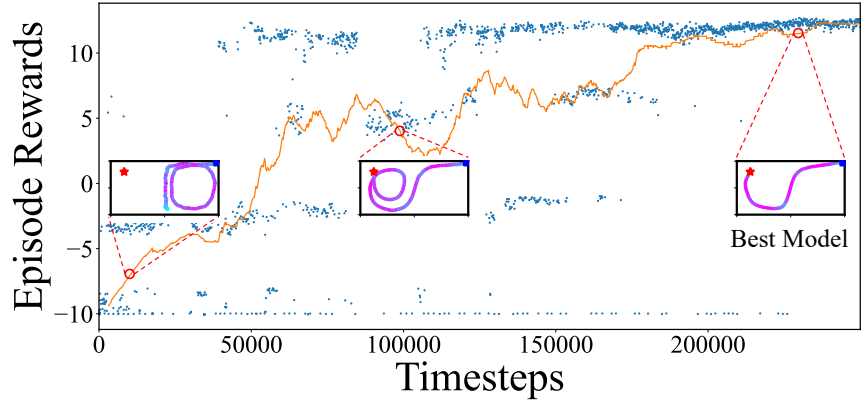}
  \caption{\label{Fig-DG-training}
  The episode rewards as a function of timesteps during the training process.
  The discrete blue dots represent rewards obtained by the agents at the different episodes, and the orange line represents a smooth average of the rewards during a short-time window of 100 episodes.
  The insets include trajectories for one failure model, one successful model with moderate reward, and the successful model with the highest reward.}
\end{figure}

We finally show the episode rewards as a function of time steps during the training process.
In Fig. \ref{Fig-DG-training}, the discrete blue dots represent the accumulated rewards obtained by the agents at each episode,
and the orange line represents the smooth average of the rewards during a short-time window of 100 episodes.
A reward value around -10 indicates that the agent migrates outside of the flow domain received a penalty; a
reward value between -5 and 0 indicates that the agent neither migrates outside of the flow domain nor reaches the destination,
but it is trapped in a gyre and its trajectories form loops;
a reward value around 5 indicates that the agent reaches the destination, yet, it takes a high energy consumption and a long migration time;
a reward value around 10 indicates that the agent can successfully reach the goal with minimal energy cost and migration time.
We can see from Fig. \ref{Fig-DG-training} that, initially (i.e., for timesteps less than 50 000),
the agent performs a random policy, and it is more likely to receive a penalty.
It migrates outside of the flow domain or does not reach the destination.
During the timesteps between 50 000 and 100 000, the agent has more chances to reach the destination and receives a positive reward.
However, the agent may still fail the migration task due to additional explorations,
which is a feature of the underlying governing reinforcement learning algorithm.
For timesteps larger than 100 000, the orange line gradually reaches a plateau with positive rewards, suggesting the training process converges.
In Fig. \ref{Fig-DG-training}, we also include trajectories for one failure model (with a reward around -5)
and one successful model with moderate reward (around 5).
For the failure model, we can see the agent only oscillates inside the right gyre, and its trajectory forms periodic orbits.
For the successful model with moderate reward, the agent bypass yet misses the destination for the first time.
It can then drift with the background flow and reach the destination for the second time.

\begin{figure}
  \centering
  \includegraphics[width=17cm]{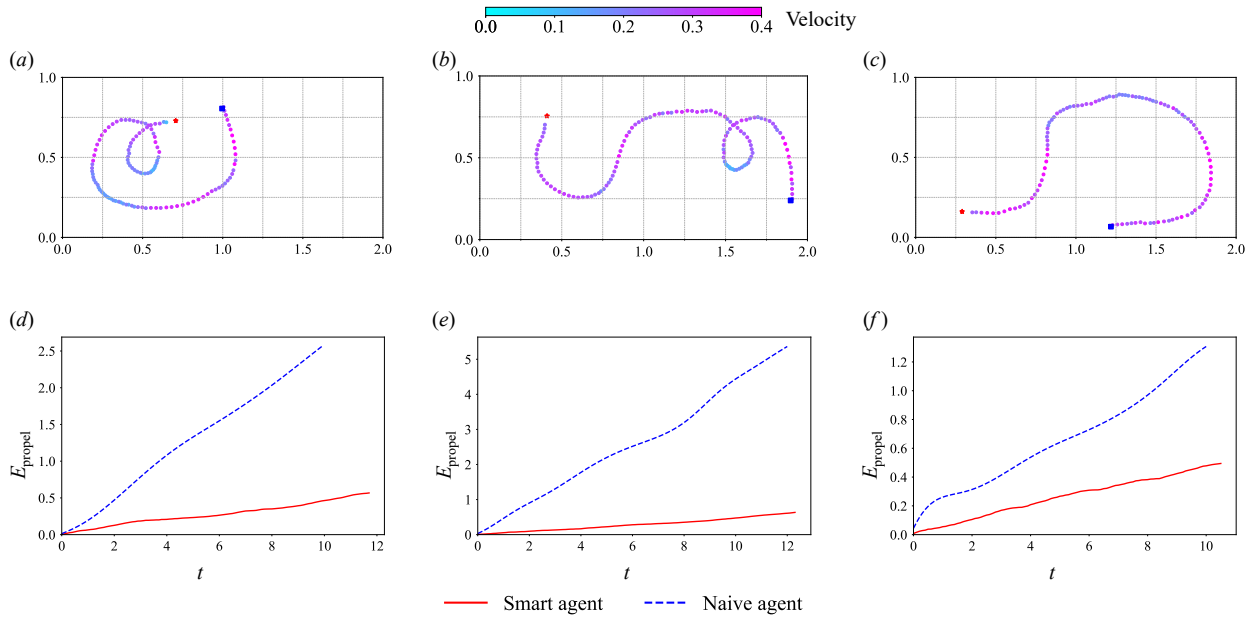}
  \caption{\label{Fig-traj_and_ecompare}
  (\textit{a})-(\textit{c}) Trajectories for the smart agent in the double-gyre flow and (\textit{d})-(\textit{f}) and the corresponding accumulative energy consumption compared to that of the naive agent.
  }
\end{figure}

The above results are obtained with the prescribed and fixed origin and destination positions for the agent.
We further test the robustness of the energy-efficient strategies with respect to random positions of the origin and the destination.
We choose the origin position in the right-half of the domain (i.e., $1 < x < 2$ and $0 < y < 1$) and the destination position in the left-half of the domain (i.e., $0 < x < 1$ and $0 < y < 1$).
We performed 100 experiments with various origin and destination positions, and three typical optimized trajectories are shown in Figs. \ref{Fig-traj_and_ecompare}(a)-\ref{Fig-traj_and_ecompare}(c).
We can see that the smart agents always try to utilize the current of the carrier flow as much as possible.
In addition, we calculate their accumulative energy consumption compared to that of the naive agents, and we plot the results in Figs.
\ref{Fig-traj_and_ecompare}(d)-\ref{Fig-traj_and_ecompare}(f).
We can see that the smart agents still consume far less energy compared to the naive agents with the randomly chosen origin and destination positions.

\section{\label{Section4}Migration in the turbulent Rayleigh-B\'enard convection}

\subsection{Numerical simulation of the turbulent Rayleigh-B\'enard convection}
We consider an incompressible thermal flow in the Oberbeck-Boussinesq approximation.
The temperature is treated as an active scalar, and its influence on the velocity field is realized through the buoyancy term.
The governing equations read as \cite{xia2013current}
\begin{subequations}
\begin{align}
& \nabla \cdot \mathbf{u}=0 \\
& \frac{\partial \mathbf{u}}{\partial t}+\mathbf{u}\cdot \nabla \mathbf{u}=-\frac{1}{\rho_{0}}\nabla P+\nu \nabla^{2}\mathbf{u}+g\beta_{T}(T-T_{0})\hat{\mathbf{y}} \\
& \frac{\partial T}{\partial t}+\mathbf{u}\cdot \nabla T=\alpha_{T} \nabla^{2} T
\end{align} \label{Eq.NS}
\end{subequations}
where $\mathbf{u}=(u,v)$, $P$ and $T$ are velocity, pressure, and temperature of the fluid, respectively.
$\rho_{0}$ and $T_{0}$ are the reference density and temperature, respectively.
$\hat{\mathbf{y}}$ is the unit vector parallel to the gravity.
$g$ is the gravity acceleration value.
$\nu$, $\beta_{T}$, $\nu$, and $\alpha_{T}$ are the kinetic viscosity, thermal expansion coefficient, kinematic viscosity, and thermal diffusivity of the fluid, respectively.
We adopt the lattice Boltzmann (LB) method as the numerical tool to solve the above equations.
The advantages of the LB method include easy implementation and parallelization.  \cite{xu2017lattice,chen1998lattice,aidun2010lattice}
More numerical details on the LB method and validation of the in-house code can be found in our previous work.  \cite{xu2017accelerated,xu2019lattice}
In simulation, the top and bottom walls of the convection cell are kept at constant
cold temperature  $T_{\text{cold}}$ and hot temperature  $T_{\text{hot}}$, respectively;
while the other two vertical walls are adiabatic.
All four walls impose no-slip velocity boundary conditions.
The dimension of the cell is $L \times H$, and we set $L=2H$ in this work.
Simulation results are provided for the Prandtl number of $Pr = \nu/\alpha = 0.71$ and
the Rayleigh number of $Ra = g\beta_{T}(T_{\text{hot}}-T_{\text{cold}})H^{3}/(\nu \alpha)  = 10^{8}$.
The $Pr$ corresponds to the thermal properties of air,
while the $Ra$ is far less than that in the atmosphere due to the limitation of computational resources to simulate ultra-high Ra convection.
Nevertheless, we can observe the large-scale coherent structure consisting of two primary rolls horizontally stacked in the simulation domain. \cite{zhou2021large,xu2020correlation,xu2022production}

\subsection{Training results and discussion}
\begin{figure}
  \centering
  \includegraphics[width=12cm]{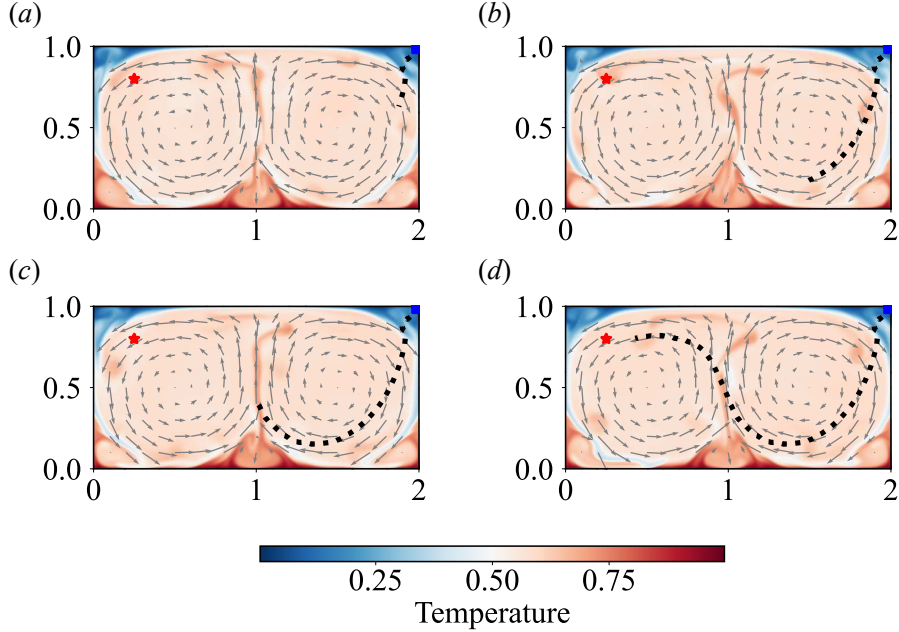}
  \caption{\label{Fig-RB_trajectory}
Trajectory (black dotted line) of the smart agent in the two-dimensional turbulent Rayleigh-B\'enard convection at
(\textit{a}) $t = 15$, (\textit{b}) $t = 30$, (\textit{c}) $t = 45$, and (\textit{d}) $t = 65$.
The contour shows the temperature field, and the vectors denote the velocity field of the convection. (Multimedia view)}
\end{figure}

In training, we restrict the maximum propelling velocity of the agent along either $x$- or $y$-direction to be less than 0.02,
which leads to  $\left\|\mathbf{u}_{\text{propel}}\right\|^{2}\le 0.0008$.
We show the instantaneous trajectories for the smart agent in the turbulent Rayleigh-B\'enard convection in Fig. \ref{Fig-RB_trajectory} (Multimedia view).
We released the agent at (2.0, 1.0), namely, the top-right corner in the domain (marked by the blue square in the plot).
The agent's goal is to reach the destination position of (0.25, 0.8), marked by the red star in the plot
with minimal energy consumption.
Initially, the smart agent will move downward to utilize the vertical currents in the top-right region [see Fig. \ref{Fig-RB_trajectory}(a)].
When the agent reaches the bottom-right corner of the right roll, it will move leftward to utilize the horizontal currents [see Fig. \ref{Fig-RB_trajectory}(b)].
After that, it will drift into the bottom-right corner of the left roll [see Fig. \ref{Fig-RB_trajectory}(c)]
and then follow the vertical upward current to achieve thermal soaring.
When the agent reaches the same altitude as the destination, it will move leftward to utilize the horizontal current [see Fig. \ref{Fig-RB_trajectory}(d)].
The smart agent tries to follow the background currents as much as possible,
which is similar to that in the double-gyre flow.
In addition, the correlation coefficient between the orientation of the propelling velocity vector (i.e., $\theta_{\text{propel}}$)
and the orientation of the fluid velocity vector (i.e.,  $\theta_{\text{fluid}}$) is 0.59, suggesting the statistical relevance between them.

\begin{figure}
  \centering
  \includegraphics[width=12cm]{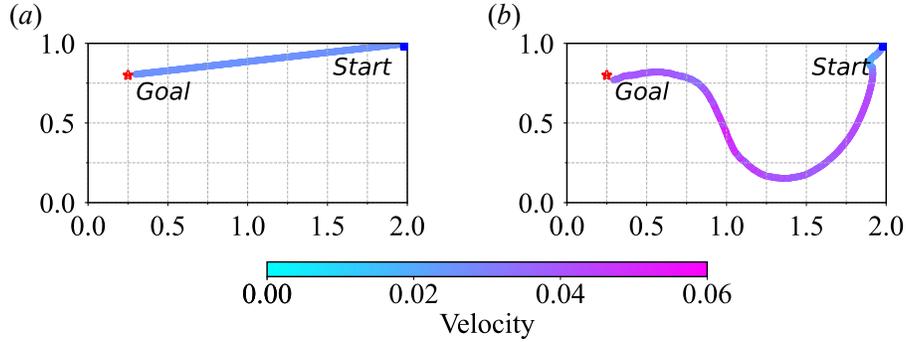}
  \caption{\label{Fig-RB_trajectory-compare}
Comparison of trajectories in the Rayleigh-B\'enard convection for (\textit{a}) a naive agent moving straightly
and (\textit{b}) a smart agent utilizing the flow structure to save energy.
The trajectories are color-coded by the instantaneous velocity magnitude. }
\end{figure}

In the turbulent Rayleigh-B\'enard convection, we also compare the smart agent with the naive agent that moves straight from the origin to the destination.
In Fig. \ref{Fig-RB_trajectory-compare}, we show naive and smart agents' trajectories, which are color-coded by the instantaneous velocity magnitude.
We set the naive agent spending the same time as the smart agent to migrate from the origin to the destination.
The naive agent migrates slower than the smart agent, because it travels a shorter distance than the smart agent.
The trajectories differences for the smart agent in the double-gyre flow (see Fig \ref{Fig-DG_trajectory})
and the turbulent Rayleigh-B\'enard convection (see Fig \ref{Fig-RB_trajectory}) are mainly due to the rotational direction of the primary vortex in the flows,
while the underlying principles to utilize the flow structure to save energy remain the same for the smart agent.

\begin{figure}
  \centering
  \includegraphics[width=12cm]{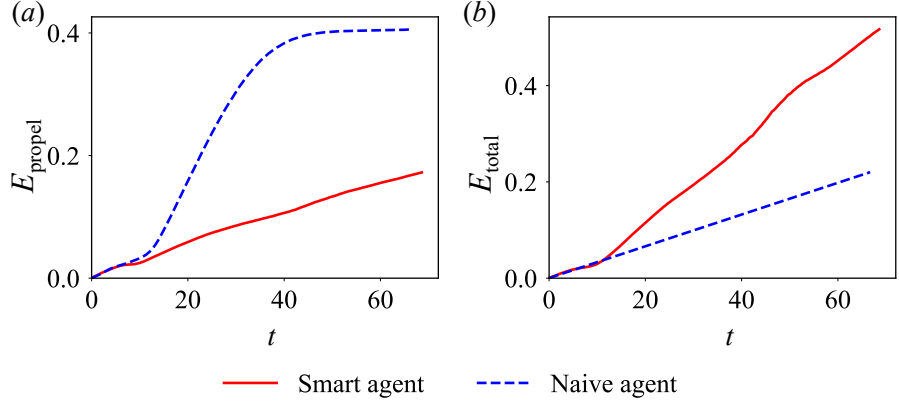}
  \caption{\label{Fig-RB-energy-compare}
Comparison of the accumulative (\textit{a}) energy consumption $E_{\text{propel}}$ and (\textit{b}) total kinetic energy $E_{\text{total}}$ of the smart agent and the naive agent in the Rayleigh-B\'enard convection.}
\end{figure}

We then plot the time series of accumulative energy consumed by the naive and smart agents in Fig. \ref{Fig-RB-energy-compare}(a).
We can see that both agents consume almost the same amount of energy during the initial period (i.e., around  $t<10$).
The reason is that the background flow velocity is relatively small [i.e., around $O(10^{-5})$ near the origin position],
and the migration of both agents heavily relies on their own propelling energy.
After that (i.e., around  $t>10$), the smart agent moves downward to utilize the flow currents,
while the naive agent continues to move straight toward the destination.
Migrating reversely against the flow currents and crossing the edge of the roll will require substantial energy consumption,
as evident from the significant increase in energy consumption at  $t>10$ [see Fig. \ref{Fig-RB-energy-compare}(a)].
The results suggest that in turbulent flows with strong background flow velocity fluctuations,
the smart agent can still save energy consumption while migrating to the destination.
We also compare the accumulative total kinetic energy of the agents [see Fig. \ref{Fig-RB-energy-compare}(b)].
After reaching the destination, the total kinetic energy of the smart agent is triple as that of the naive agent,
while the smart agent consumed almost one-third of the energy.

\begin{figure}
  \centering
  \includegraphics[width=12cm]{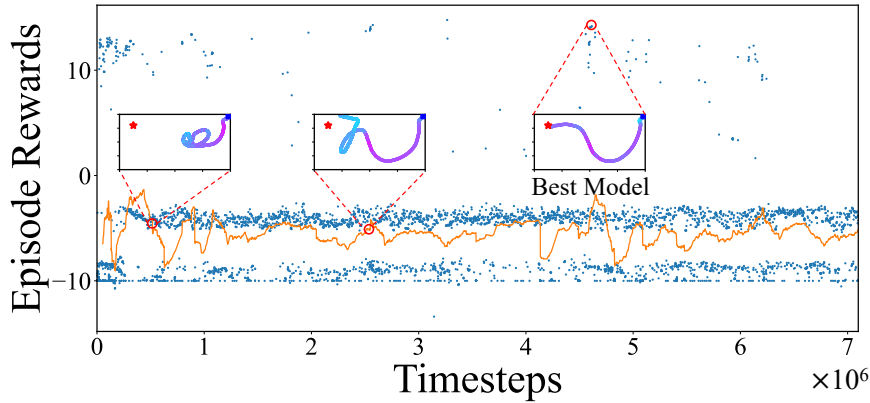}
  \caption{\label{Fig-RB-training}
The episode rewards as a function of timesteps during the training process.
The discrete blue dots represent rewards obtained by the agents at the different episodes,
and the orange line represents a smooth average of the rewards during a short-time window of 100 episodes.
The insets include trajectories for two failure models and the successful model with the highest reward.}
\end{figure}

We show the episode rewards as a function of time steps during the training process in Fig. \ref{Fig-RB-training}.
The discrete blue dots represent rewards obtained by the agents at the different episodes.
A reward value around -10 indicates that the agent migrates outside of the flow domain received a penalty;
a reward value between -5 and 0 indicates that the agent neither migrates outside of the flow domain nor reaches the destination
but is trapped in the right roll and oscillates inside the right roll;
a reward value around 10 indicates that the agent can successfully reach the goal with minimal energy cost and migration time.
We can see from Fig. \ref{Fig-RB-training} that the training processes for the agent in the turbulent flows can hardly converge
in contrast to that in the simple periodic double-gyre flow.
The main reason is that the flow field in the turbulent Rayleigh-B\'enard convection exhibits more substantial fluctuations,
and the migrating agent controlled by the reinforcement learning algorithm is more likely to explore different migration strategies in each episode.
We determine the policy $\pi^{*}(\theta)$ that associated with the highest reward as the optimal policy, and we then train the agent with $\pi^{*}(\theta)$  to obtain the trajectory corresponding to the "best model".
Due to the fluctuations of the flow field in the turbulent Rayleigh-B\'enard convection, the trajectories generated by $\pi^{*}(\theta)$  may be slightly different in different runs.
In Fig. \ref{Fig-RB-training}, we include trajectories for two failure models (with a reward around -5 and -10, respectively)
and one successful model with the highest reward (around 10).
For the failure models, we can see the agent either oscillates in the right roll and does not drift into the left roll (around -5),
or migrates outside of the domain (around -10).

\begin{figure}
  \centering
  \includegraphics[width=12cm]{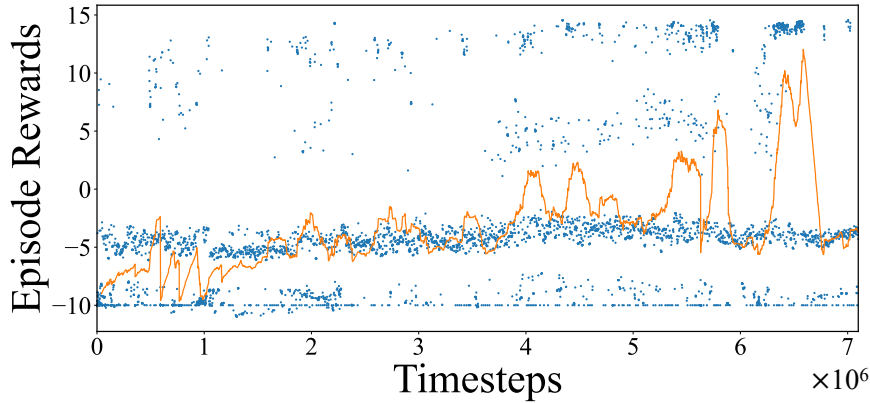}
  \caption{\label{Fig-RB-training_withT}
  The episode rewards as a function of timesteps during the training process.
  Different from the results presented in Fig. \ref{Fig-RB-training}, here, we also include temperature as sensorimotor during the training process.
}
\end{figure}

The above results are obtained with observation variables including flow velocity, the agent's spatial coordinates, and current time; however, in the turbulent Rayleigh-B\'enard convection, the temperature acts as an active scalar that influences the velocity;
thus, in Fig. \ref{Fig-RB-training_withT} we further show the episode rewards in which the fluid temperature is also considered as additional observation variables during the training process.
We can see that including temperature as sensorimotor, indeed, improves the average episode rewards during the training process, as evident there are more chances for the agent to obtain a high episode reward.
On the other hand, we checked the optimized trajectories and the associated energy consumptions of the agents, and we found that including temperature as sensorimotor only slightly changes the optimized policy.

\section{\label{Section5}Conclusions}
In this work, we performed training of the self-propelling agent to migrate in a flow environment with the reinforcement learning algorithm.
The smart agent that migrates in both the two-dimensional periodical double-gyre flow and two-dimensional turbulent Rayleigh-B\'enard convection can learn to migrate from one position to another
while utilizing the background flow currents as much as possible.
We calculated the energy consumption for a naive agent that moves straight from the origin to the destination.
The results show that the smart agent consumed less than one-fifth (or one-third) of energy compared to the naive agent in the periodical double-gyre flow (or the turbulent Rayleigh-B\'enard convection).
In addition, we found that compared to the double-gyre flow, the flow field in the turbulent Rayleigh-B\'enard convection exhibits more substantial fluctuations,
and the training agent is more likely to explore different migration strategies.
Despite that the training process is challenging to converge for the smart agent in a turbulent environment,
we can identify an energy-efficient trajectory that corresponds to the strategy with the highest reward received by the agent.
We also found that including temperature as sensorimotor improves the average episode rewards during the training process in the turbulent Rayleigh-B\'enard convection, while the optimized trajectories and the associated energy consumption of the agents almost remain unchanged.
As pointed out by Laurent \emph{et al.}, \cite{laurent2021turbulence} there are opportunities to harness the energy of turbulence, particularly for person-less transport and small reconnaissance aircraft.
Thus, similar processes could very well be optimized in other migration involving a self-propelling agent, such as UAVs flying in a turbulent convective environment.


\begin{acknowledgments}
This work was supported by the National Natural Science Foundation of China (NSFC) through grant Nos. 11902268 and 12125204,
the National Key Project via No. GJXM92579,
and the 111 project of China (No. B17037).
\end{acknowledgments}

\section*{Author Declarations}
\subsection*{Conflict of Interest}
The authors have no conflicts to disclose.

\section*{Data Availability}
The data that support the findings of this study are available from the corresponding author upon reasonable request.

\section*{Appendix: Propulsion velocity vector of the smart agent in the double-gyre flow}
\begin{figure}
  \centering
  \includegraphics[width=12cm]{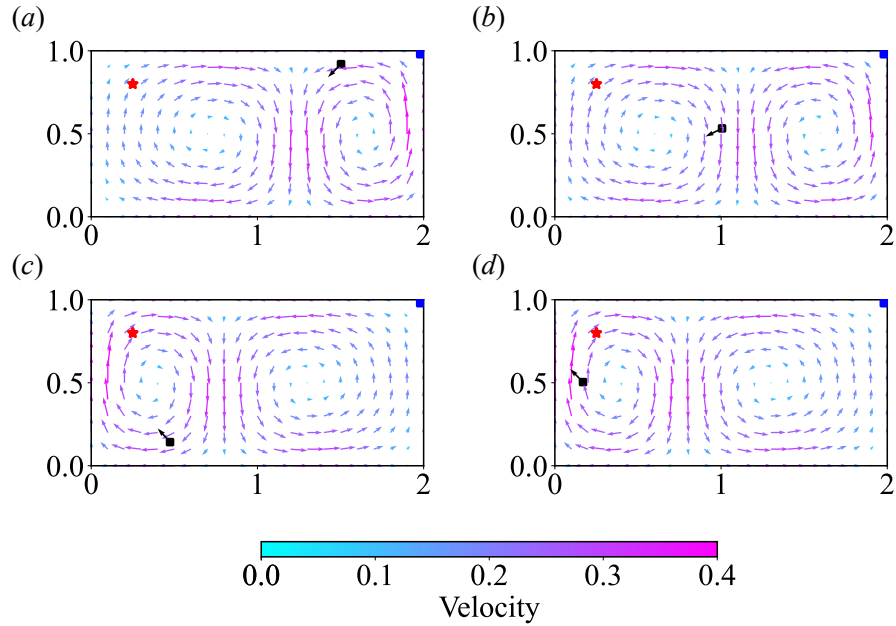}
  \caption{\label{Fig-DG_trajectory_with_prop_velocity} Propulsion velocity vector (black vector) of the smart agent in the double-gyre flow at
  (\textit{a})  $t = 2.2$, (\textit{b})  $t = 4.2$, (\textit{c})  $t = 6.7$, and (\textit{d})  $t = 8.2$.
  The color-coded vectors denote the velocity field of the double-gyre flow.}
\end{figure}

We show the propulsion velocity vector of the smart agent in different locations in Fig. \ref{Fig-DG_trajectory_with_prop_velocity},
which further clarify the dynamics of the smart agent in the double-gyre flow.
The smart agent utilizes background currents as much as possible, as evident that the angle between the propelling velocity vector (the black vector) and the fluid velocity vector (color-coded vectors) is generally less than $90^{\circ}$ at the same location.

\nocite{*}
\bibliography{myBib}

\end{document}